\documentclass[preprint,aps,pre,longbibliography]{revtex4-1}
\usepackage[dvips]{graphicx}
\usepackage{fancybox}
\usepackage{amsmath,amssymb,bm,amsthm}
\def\beq{\begin{equation}}
\def\eeq{\end{equation}}

\def\<{\langle}
\def\>{\rangle}

\begin{document}

%\title{Fluctuations in a kinetic Ising model driven by a slowly oscillating 
%field with a constant bias}
\title{Fluctuations in a model ferromagnetic film driven by a slowly oscillating 
field with a constant bias}

\author{Gloria M.\ Buend{\'\i}a$^1$ and Per Arne Rikvold$^2$}
\affiliation{$^1$Department of Physics, Universidad Sim{\'o}n Bol{\'{\i}}var, 
Caracas 1080, Venezuela\\
$^2$ Department of Physics, Florida State University, 
Tallahassee, FL 32306-4350,USA
}

\begin{abstract}
We present a numerical and theoretical study that supports and explains recent experimental 
results on anomalous magnetization fluctuations of a 
uniaxial ferromagnetic film in its low-temperature phase, which is 
forced by an oscillating field above the critical period of the associated 
dynamic phase transition (DPT) 
[P.\ Riego, P.\ Vavassori, A.\ Berger, Phys.\ Rev.\ Lett.\ {\bf 118}, 117202 (2017)]. 
For this purpose, we perform kinetic Monte Carlo 
simulations of a two-dimensional Ising model with nearest-neighbor ferromagnetic interactions 
in the presence of a sinusoidally 
oscillating field, to which is added a constant bias field. 
We study a large range of system sizes and supercritical periods and analyze the data using a  
droplet-theoretical description of magnetization switching. 
We find that the period-averaged magnetization, which plays the role of the order parameter 
for the DPT, presents large 
fluctuations that give rise to well-defined peaks in its scaled variance and its 
susceptibility with respect to the bias field. The peaks are symmetric with 
respect to zero bias and located at values of the bias field that 
increase toward the field amplitude as an inverse logarithm of the 
field oscillation period. 
Our results indicate that this effect is independent of the system size 
for large systems, ruling out critical behavior associated with a phase transition. 
Rather, it is a stochastic-resonance phenomenon that has no counterpart in the 
corresponding thermodynamic phase transition, providing a reminder that the equivalence 
of the DPT to an equilibrium phase transition is limited to the critical region near the 
critical period and zero bias.
\end{abstract}

%\pacs{75.30.Wx,64.60.My,64.60.Kw,75.60.-d}

\maketitle

%*******************************************************************************
%\begin{document}

\section{Introduction}
\label{sec:I}

The hysteretic response when a uniaxial spin system with long-range order 
({\it i.e.}, below its critical temperature) 
%, such as a kinetic Ising ferromagnet below its critical temperature, 
is subject to a symmetrically oscillating field of amplitude $H_0$ and period $P$, 
depends crucially on $P$. If $P$ is much 
longer than the response time of the system (which depends on the 
temperature and $H_0$), a symmetric hysteresis loop centered on zero results. 
If $P$ is much shorter than the response time, asymmetric hysteresis loops 
centered around the values of the system's static order parameter are observed. 
Numerical studies in the 1990's showed that the transition between these 
two regimes is not smooth. Rather, there is a critical period $P_c$, where the 
period-averaged order parameter $\langle Q \rangle$ 
(see formal definition in Sec.~\ref{sec:H}) 
vanishes in a singular fashion. This phenomenon was first observed by 
Tom{\'e} and de~Oliveira \cite{TOME90} in a kinetic mean-field study of an Ising model, 
followed by kinetic Monte Carlo (MC) simulations by Rao, Krishnamurthy, and Pandit \cite{RAO90B} 
and Lo and Pelcovitz \cite{LO90}.  
Early work in the field was reviewed 
by Chakrabarti and Acharyya  
in Ref.\ \cite{CHAK99}.
Kinetic MC combined with finite-size scaling analysis
 \cite{SIDE98,SIDE99,KORN00,ROBB07,BUEN08,PARK13}, as well as further 
mean-field studies of Ising and Ginzburg-Landau models 
\cite{FUJI01,GALL12,IDIG12,ROBB14}, confirmed not only that this is 
a true, dynamic phase transition (DPT), but also that it is in the same 
universality class as the corresponding equilibrium Ising model. The DPT has 
been confirmed experimentally in [Co/Pt]$_3$ magnetic multilayers 
\cite{ROBB08} and uniaxial Co films \cite{BERG13}.

With all the attention that has been given to the DPT and its universality 
class, one might lose sight of the fact that the equivalence between the 
critical properties of the equilibrium Ising model and the DPT of the same 
model in an oscillating field does not necessarily amount to equivalence {\it 
outside\/} the critical region. A warning was provided very recently by Riego, 
Vavassori, and Berger \cite{RIEG17}. These authors fabricated Co films 
with $(10\underbar{1}0)$ crystallographic surface structure with a single, 
in-plane magnetic easy axis, which they subjected to a sinusoidally oscillating, 
in-plane magnetic field plus a constant bias field $H_b$. Such a constant bias 
field has previously been shown by MC simulations and finite-size scaling to be 
(at least a significant component of) the field conjugate to $\langle Q \rangle$ in the 
critical region near $P_c$ \cite{ROBB07}, and this has later been confirmed for 
mean-field models \cite{GALL12,IDIG12,ROBB14} and in experiments \cite{BERG13}. 
It therefore seemed 
surprising that, in the experiments reported in Ref.\ \cite{RIEG17}, 
both the fluctuations in the order parameter and its derivative with respect to 
$H_b$, for $P \gg P_c$, behaved quite differently from the dependence of the 
equilibrium susceptibility on the applied static field at temperatures above 
critical. Instead of the wide, smooth, unimodal maximum of the supercritical equilibrium 
susceptibility of the Ising model, 
two distinct peaks were observed at nonzero values of $H_b$, symmetrical 
about zero  \cite{RIEG17}. In their article the
authors also presented kinetic mean-field results that corroborate the presence of 
these peaks, which they dubbed ``sidebands."

The purpose of the present paper is to investigate the long-period parameter 
regime with kinetic MC simulations of a two-dimensional Ising model with 
nearest-neighbor ferromagnetic interactions. 
To match the experimental conditions of Ref.\ \cite{RIEG17} as closely as possible, we 
choose the oscillating field to have a sinusoidal waveform. 
We are not aware that systematic simulations in this regime have been performed previously. 
Our study reveals ``sidebands" analogous to the experimental results. 
We thus conclusively confirm that the experimentally observed phenomenon is not caused by 
residual magnetostatic long-range interactions. 
Using simulations for a range of field periods and system sizes 
together with knowledge of the kinetics of magnetization switching by homogeneous 
nucleation and growth of antiphase droplets \cite{RIKV94A}, 
we demonstrate that the ``sidebands"  result from noncritical fluctuations during the 
half-cycles when the sign of the oscillating field is opposite to that of the bias field. 
This is essentially a stochastic resonance phenomenon \cite{GAMM98,SIDE98A,KORN02B}. 

The rest of this paper is organized as follows. 
In Sec.~\ref{sec:H} we describe the model and details of the simulation method, and we define 
the appropriate observables to be measured. 
Our numerical results are presented in Sec.~\ref{sec:R}. 
In Sec.~\ref{sec:SB} 
we present numerical observation of sidebands for a single, supercritical 
value of the field period. 
In Sec.~\ref{sec:mt} we present short time series of 
the system magnetization for several values of bias and period, 
which enable us to propose a simple approximation for $\langle Q \rangle$ in the limits of 
weak bias and long period. In Sec.~\ref{sec:HP} we present numerical results for 
$\langle Q \rangle$
%$\chi_L^Q$, and $\chi_L^b$ 
vs $H_b$ for a wide range of supercritical periods, as well 
as the sideband positions $H_b^{\rm peak}$ as functions of period and system size. 
The latter are analyzed using results from the droplet theory of magnetization reversal. 
Our conclusions are given in Sec.~\ref{sec:CONC}. A short summary  
of pertinent results from the droplet theory of magnetization reversal is given in  
Appendix \ref{sec:mech}, and the case of extremely long periods is discussed in Appendix 
\ref{sec:CAV}. A brief discussion of the mathematically 
simpler case of a square-wave oscillating field is presented in Appendix \ref{sec:SQW}.

\section{Model and Monte Carlo Simulation}
\label{sec:H}
We consider a kinetic $S=1/2$ Ising model with a time-dependent external 
field and ferromagnetic nearest-neighbor interactions.  Its Hamiltonian is 
\begin{equation}
{\mathcal H}=-J\sum_{\langle ij \rangle} s_{i}s_{j} - \left[ H(t)+H_{b} \right] \sum_{i} s_{i} \;,
\label{eq=ham}
\end{equation}
where $J>0$, $s_{i}=\pm1$, the first sum runs over all nearest-neighbor pairs, 
and the second one over all sites. $H_b$ is a constant ``bias field," and $H(t)$ 
is a symmetrically oscillating external field of period $P$. Here we choose
\begin{equation}
 H(t)=H_{0}  \cos \left( \frac{2 \pi}{P} t \right) \;.
\label{eq:field}
\end{equation}
The system is simulated on a square lattice of $N=L \times L$ sites with 
periodic boundary conditions.  
We perform Glauber single-spin-flip dynamics in a heat bath at  
temperature $T$. A spin at a randomly chosen site $i$ is allowed to flip from 
$s_i$ to $-s_i$ with probability 
\begin{equation}
 W(s_{i}\rightarrow -s_{i})=\frac {1}{1+\exp(\beta \Delta E_{i})} \;,
\label{eq:W}
\end{equation}
where $\Delta E_{i}$ is the change in the system energy associated 
with flipping the spin $i$, and $\beta=1/k_{\rm B}T$ where $k_{\rm B}$ is 
Boltzmann's constant. The time unit is one MC step per site (MCSS), 
during which, on average, each site is visited once. Hereafter, $H_0$, $H_{b}$, and $T$ are 
all given in units of the interaction constant $J$ (i.e., $J=k_{\rm B}=1$), 
and $P$ is given in units of MCSS. 

The Glauber dynamic can be derived as 
the weak-coupling limit of the quantum-mechanical Hamiltonian of a collection 
of quasi-free Fermi fields in thermal equilibrium with a heat bath 
\cite{MART77}. However, the DPT with $H_{b}=0$ has been shown to be universal 
with respect to dynamics that obey detailed balance in equilibrium, including 
Metropolis \cite{VATA17} and ``soft Glauber" \cite{BUEN08}, as well as different 
forms of $H(t)$ including square-wave \cite{KORN00,BUEN08} and sawtooth 
\cite{ROBB08}.

We calculate the time dependent, normalized magnetization per site,
\begin{equation}
 m(t)=\frac{1}{L^2} \sum_{i}s_{i}(t) \;,
\label{eq:mag}
\end{equation}
and by integrating it over each cycle of the 
magnetic field, we obtain the average magnetization during the $k$th 
cycle of the field,
\begin{equation}
 Q_{k}=\frac{1}{P}  \int_{(k-1)P}^{kP} m(t)dt \;.
 \label{eq:q}
\end{equation}
The dynamic order parameter of the model 
is the period-averaged magnetization, $\langle Q \rangle$, 
defined as the average of $Q_k$ over many cycles. Its fluctuations are measured by the scaled variance,
\begin{equation}
\chi_{L}^{Q}=L^{2}(\langle Q^{2} \rangle -\langle Q \rangle^{2}) \;,
\label{eq:var}
\end{equation}
and its dependence on the bias field is measured by the susceptibility with respect to $H_{b}$,
\begin{equation}
\chi_{L}^{b}=d\langle Q \rangle/dH_{b} \;.
\label{eq:sus}
\end{equation}

In order to take advantage of temperature and field dependent parameters measured 
with high precision in previous MC simulations \cite{SIDE99}, 
our calculations are performed with $H_{0}=0.3$ at $T=0.8T_{c}$, where 
$T_{c}={2}/{\ln(1+\sqrt {2})} \approx 2.269$ is the critical temperature of the standard, 
square-lattice Ising model in zero field. In the absence of a bias field, at this 
temperature, and for sufficiently large $L$, 
switching between the equilibrium values of $m$, following field reversal from 
$-H_0$ to $+H_0$, occurs via a 
nearly deterministic and $L$-independent multi-droplet mechanism \cite{RIKV94A}. 
In Ref.\ \cite{SIDE99}, the characteristic switching timescale (the time from the field 
reversal until the system magnetization reaches zero) under Glauber dynamics with the same 
parameters as we use here was measured by 
MC simulations as $\tau_0 \approx 74.6$. 
In the same work, the critical period in a sinusoidal field of amplitude $H_0$ with zero bias 
was measured as $P_c \approx 258$. 

The cycle-averaged magnetization 
$\langle Q \rangle$ vanishes for $P \ge P_c$ and $H_b=0$. 
Near criticality, the constant bias field $H_{b}$ is the field conjugate to $\langle Q \rangle$, and 
the period $P$ mimics the temperature in the equilibrium phase transition. 
Simulations were performed for periods between $P = 258$ and 28,000 and 
system sizes between $L = 32$ and 1024.  
Except for the smallest values of $P$, the measurements were obtained by averaging 
over 800 field cycles, after discarding 200 cycles. This means that 
at least $800 \times P$ MCSS were performed for each measurement.

\section{Numerical Results and Analysis}
\label{sec:R}

\subsection{Observation of ``sidebands"}
\label{sec:SB}
\begin{figure}[ht]
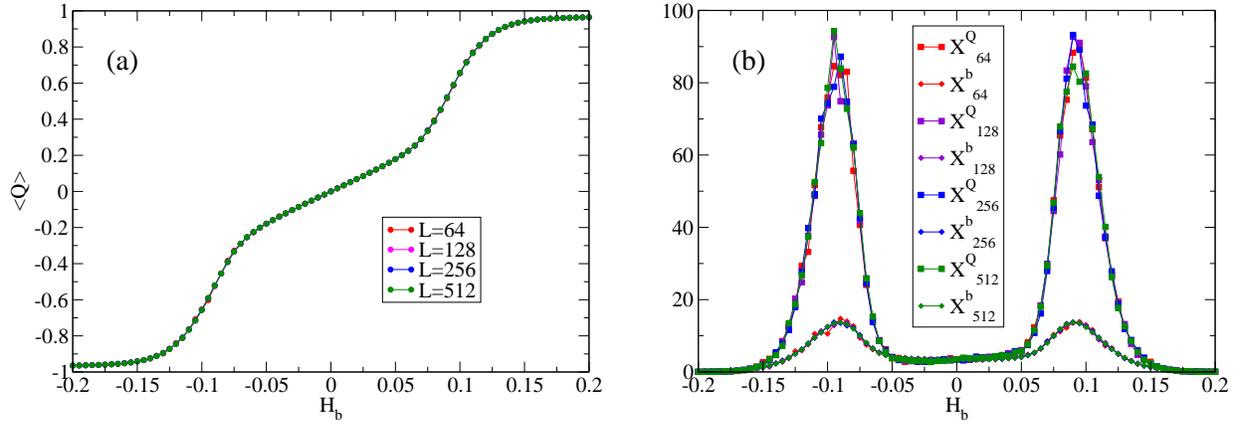

\begin{center}
\vspace*{-0.8truecm}
\includegraphics[angle=0,width=.48\textwidth]{tests_Q_cosX.eps} 
\hspace*{0.5truecm}
\includegraphics[angle=0,width=.46\textwidth]{tests_Chi_cosX.eps} 
\end{center}
\vspace*{-0.3truecm}
\caption[]{
%\baselineskip=0.3truecm
\baselineskip=0.15truecm
Results with $P = 1000 \approx 3.9P_c$ for system sizes $L=64$, 128, 256, and 512. 
With this period length and range of system sizes, finite-size effects are negligible, and the 
curves for different system sizes practically coincide.
(a)
The order parameter $\langle Q \rangle$ vs $H_b$. Error bars are smaller than the symbol size. 
(b) 
The scaled variance $\chi_{L}^{Q}$ and susceptibility $\chi_{L}^{b}$. 
%Both are seen to vanish for strong bias, have two peaks at $H_{b} \approx \pm 0.09$, and then a 
%somewhat reduced, but still significant value for $H_{b} \approx 0$. 
See discussion of this figure in Sec.\ \protect\ref{sec:SB}. 
}
\label{fig:size1}
\end{figure}

Results of simulations with $P = 1000 \approx 3.9P_c$ 
for several values of $L$ are displayed in 
Fig.\ \ref{fig:size1}. ``Sidebands" are observed, consistent with the 
experiments reported in Ref.\ \cite{RIEG17}. The dependence of 
the order parameter $\langle Q \rangle$ on the bias $H_b$ is shown in Fig.\ \ref{fig:size1}(a). 
For weak
$H_b$, $\langle Q \rangle$ increases almost linearly with $H_b$, but the slope of the curve
increases considerably around $|H_b| \approx 0.09$, followed by saturation of 
$\langle Q \rangle$ for $|H_b| \gtrsim 0.15$. 
This behavior is reflected in the bimodal shape of the susceptibility $\chi_L^b$, 
shown by the lower set of curves 
in Fig.\ \ref{fig:size1}(b). Between the two peaks lies a flat-bottomed valley corresponding 
to the linear regime in part (a), and a rapid approach to zero for large $|H_b|$ mirrors the 
saturation of $\langle Q \rangle$ also seen in (a). 
The scaled variance $\chi_L^Q$ also displays peaks, whose positions coincide with those of 
$\chi_L^b$. However, the ratio $\chi_L^Q / \chi_L^b$ for fixed $P$ 
depends quite strongly on $H_b$ with 
maximum values near the peaks. This variable ratio precludes a straightforward interpretation 
in terms of an effective, nonequilibrium fluctuation-dissipation relation with $P$ playing the 
role of ``temperature."
For these values of $L$ and $P$, finite-size effects are seen to be negligible, 
ruling out critical behavior associated with a phase transition. 
The relationships between system size, field period, and finite-size effects will be discussed 
in further detail below. 

\subsection{Magnetization time series}
\label{sec:mt}
\begin{figure}[ht]
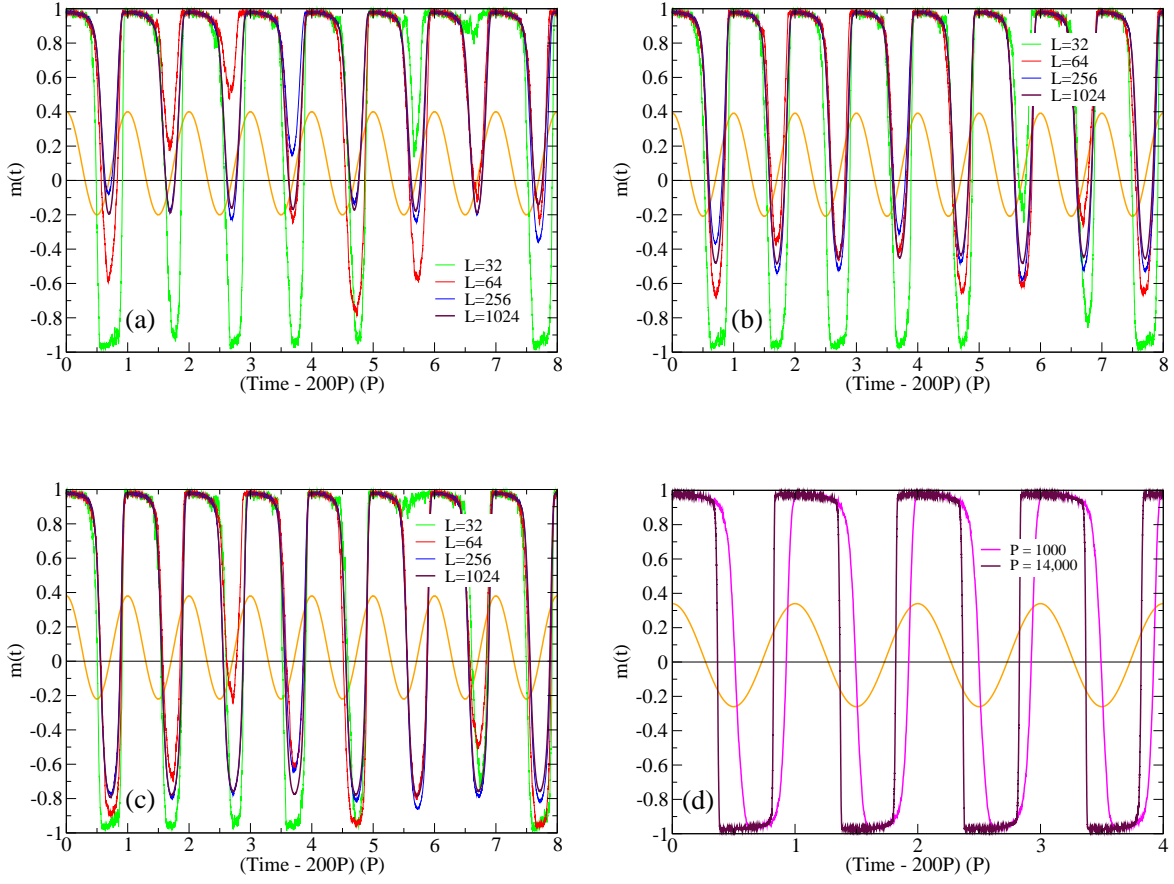

\begin{center}
\vspace*{0.8truecm}
\includegraphics[angle=0,width=.45\textwidth]{TenPeriods_P1000_Hb0_0100_VX.eps} 
\hspace*{0.4truecm}
\includegraphics[angle=0,width=.45\textwidth]{TenPeriods_P1000_Hb0_00915_VX.eps} 
\end{center}
\vspace*{0.2truecm}
\begin{center}
\includegraphics[angle=0,width=.45\textwidth]{TenPeriods_P1000_Hb0_00800_VX.eps}
\hspace*{0.4truecm}
\includegraphics[angle=0,width=.45\textwidth]{TenPeriods_Hb004_L128_VX.eps} 
\end{center}
\vspace*{-0.3truecm}
\caption[]{
%\baselineskip=0.3truecm
\baselineskip=0.15truecm
The time-dependent magnetization $m(t)$ over a few cycles following a $200P$ stabilization run, 
using systems with $L$ between 32 and 1024. 
In all four parts, the bias is positive, and the total applied field, $H(t) + H_b$, 
is shown as an orange cosine curve. 
A detailed discussion of this figure is given in Sec.\ \ref{sec:mt}. 
(a)
$P=1000$ and $H_b=+0.10$, just on the strong-bias side of the fluctuation 
peak for this period length. 
(b)
$P=1000$ and $H_b=+0.0915$, at the maximum of the fluctuation peak. 
(c)
$P=1000$ and $H_b=+0.08$, just on the weak-bias side of the fluctuation peak. 
(d)
$L=128$ and a weak bias $H_b = +0.04$ with two different period lengths, $P=1000$ 
and 14,000. The switching is deterministic and complete, and as 
$P$ increases, it occurs earlier in the 
half-period. This observation suggests the asymptotic weak-bias, long-period approximation 
for $\langle Q(H_b/H_0) \rangle$, given in 
Eq.~(\ref{eq:QQ}) and included in Fig.~\protect\ref{fig:128vsh}(a). 
}
\label{fig:tenP}
\end{figure}

To gain a more detailed understanding of the relationships between bias, period, system size, 
and the order-parameter fluctuations, we present in Fig.\ \ref{fig:tenP} short time series of the normalized magnetization, $m(t)$. 
The total applied field, $H(t) + H_b$, is shown as an orange curve. 
In this figure we set $H_b > 0$, so that the up-spin phase is favored and 
the down-spin phase is disfavored. 

Figure \ref{fig:tenP}(a) shows data for 
$P=1000$ and $H_b=+0.10$, just on the strong-bias side of the fluctuation 
peak for this period length. 
For the smaller system sizes ($L=32$ and 64), the switching from the favored 
(up-spin) to 
the disfavored (down-spin) magnetization is stochastic and abrupt 
(mediated by a single or a few droplets of the down-spin phase \cite{RIKV94A}) 
and occurs only in narrow time windows near the negative extrema of the total applied 
field. For the larger systems, the switching becomes more deterministic and gradual 
(multi-droplet \cite{RIKV94A}). 
However, the growing down-spin phase does not have time 
to completely fill the system before the field 
again becomes positive. For the largest system studied, $L=1024$, the extreme 
negative magnetizations during a period are close to $-0.2$.

Figure \ref{fig:tenP}(b) shows data for 
$P=1000$ and $H_b=+0.0915$, at the maximum of the fluctuation peak. 
The switching behavior for $L=32$ 
remains stochastic. However, the larger systems appear more deterministic, and their extreme 
negative magnetizations during a period are close to $-0.4$. 

Figure \ref{fig:tenP}(c) shows data for 
$P=1000$ and $H_b=+0.08$, just on the weak-bias side of the fluctuation peak. 
The switching for $L=32$ remains stochastic. The larger systems behave more deterministically, 
and the extreme negative magnetizations during a period approach $-0.8$. 

These results illustrate how the switching behavior in the peak region crosses over from 
a stochastic single-droplet mechanism for small $L$ to a nearly deterministic multidroplet 
mechanism for larger $L$, in agreement with known results for field-driven magnetization 
switching by homogeneous nucleation and growth of droplets of the stable phase \cite{RIKV94A}. 

Figure \ref{fig:tenP}(d) shows data for 
$L=128$ with a weak bias, $H_b = +0.04$, and two different period lengths, $P=1000$ 
and 14,000. In both cases, the switching is nearly deterministic and complete, so that the 
period-averaged magnetization $\langle Q \rangle$ depends mostly on the 
relative amounts of time the 
system spends in the two phases. As $P$ increases, the switching occurs earlier in the half-period. 

\begin{figure}[ht]
%\vspace*{0.8truecm}
\begin{center}
\includegraphics[angle=0,width=.7\textwidth]{NPeriods_L32_1024.eps}
\end{center}
\begin{center}
\vspace*{1.0truecm}
\includegraphics[angle=0,width=.45\textwidth]{snapL32mp01.eps} 
\hspace*{1.0truecm}
\includegraphics[angle=0,width=.44\textwidth]{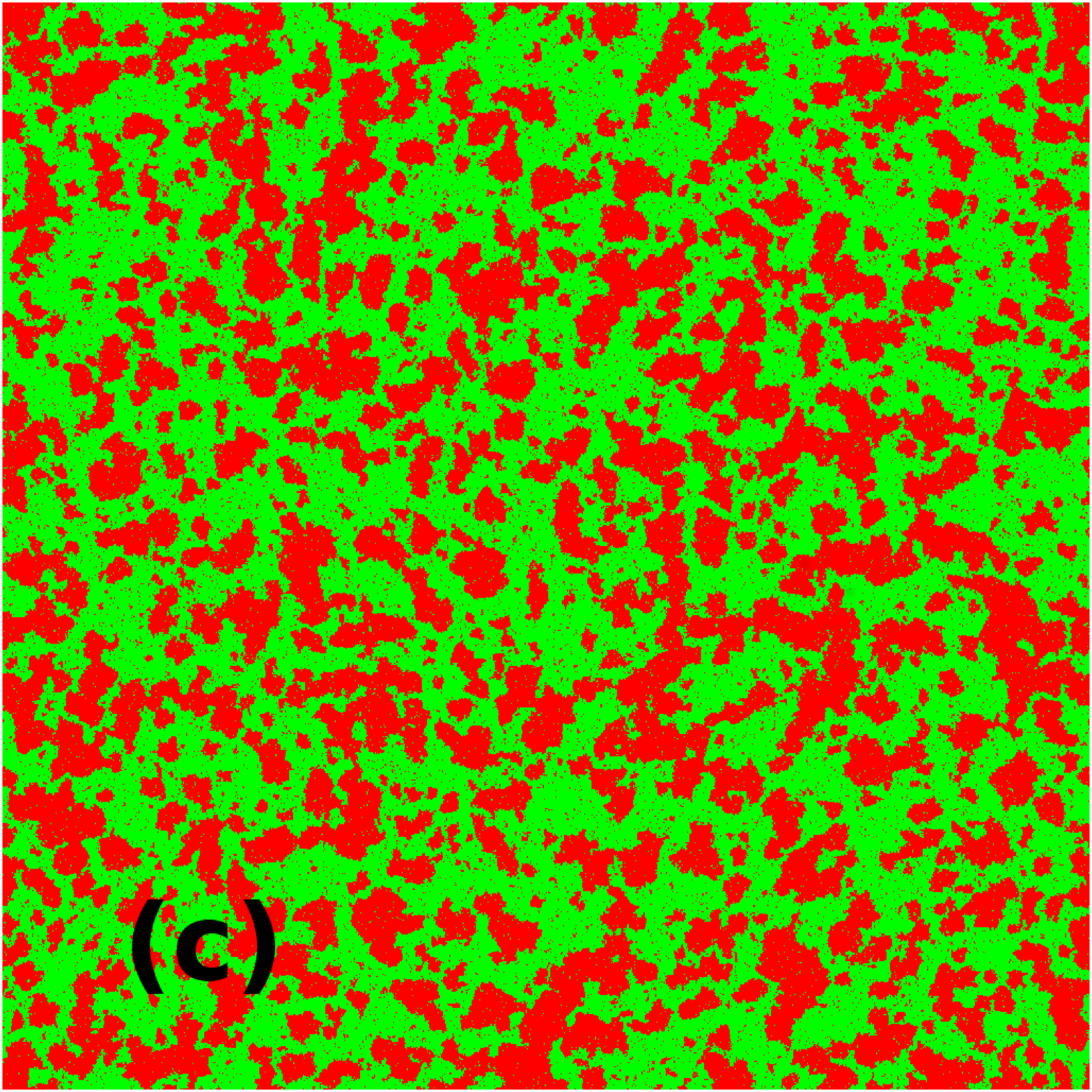} 
\end{center}
\vspace*{-0.3truecm}
\caption[]{
%\baselineskip=0.3truecm
\baselineskip=0.15truecm
A short time series and snapshots showing growing disfavored-phase clusters for $P=1000$ 
at the corresponding peak position, $H_b^{\rm peak} = +0.0915$. 
(a)
Time series $m(t)$ over five cycles following a $200P$ stabilization run, 
showing data for $L=32$ (green) and 1024 (maroon).  
The total applied field, $H(t) + H_b$, is shown as an orange cosine curve.
The snapshots were captured the first time past $200P$ that 
$m(t)$ fell below $+0.1$ (red horizontal line in the figure), 
corresponding to a disfavored-phase (down-spin) fraction of $0.45$. The times of capture are 
marked by black circles. 
In the following snapshots, regions of the up-spin phase are green, 
and down-spin are red. 
(b)
$L=32$. A single droplet of the down-spin phase has nucleated near the time when the total applied 
field has its largest negative value. The highly stochastic nature of the single-droplet switching mode 
is also evident from the time series in part (a). 
(c)
$L=1024$. Many droplets of the down-spin phase have nucleated 
at different times and then grown almost 
independently. At the moment of capture, some clusters have coalesced while others are still 
growing independently. From the time series in part (a) it is seen that this multi-droplet 
switching mode leads to a nearly deterministic evolution of the total magnetization.
This figure is further discussed in Sec.\ \ref{sec:mt}. 
}
\label{fig:SNAP}
\end{figure}

The differences between the single-droplet and multidroplet switching 
modes are further illustrated in Fig.~\ref{fig:SNAP}.
In Fig.~\ref{fig:SNAP}(a),
time series for $m(t)$ over five cycles with $P=1000$ 
at the corresponding peak position, $H_b^{\rm peak} = +0.0915$ 
show data for $L=32$ and 1024. All the parameters are the same as in 
Fig.\ \ref{fig:tenP}(b), except the seed for the random number generator.
When the total applied field, $H(t) + H_b$, is negative, 
the down-spin phase, which is 
disfavored by the positive bias, is the equilibrium phase. 
Nucleation and growth of this phase may only occur during the time intervals of 
negative total applied field. 
Snapshots captured at $m(t) = +0.1$ during these growth periods, 
corresponding to a down-spin fraction of $0.45$, are shown in 
Fig.~\ref{fig:SNAP}(b) for $L=32$ and in Fig.~\ref{fig:SNAP}(c) for 
$L=1024$. 

For $L=32$ we see a single down-spin droplet which, as seen from 
the time series in Fig.~\ref{fig:SNAP}(a), nucleated during the 
third period shown, near the time when the 
field had its largest negative value. It barely reached the capture threshold 
of $m=+0.1$ before the field again became positive and caused it to 
decay. The stochastic nature 
of this single-droplet switching mode is also 
clearly reflected by the time series. 
During the five periods shown, the capture threshold was only reached twice. 
And only once, during the fifth period, do we see full saturation of the 
down-spin phase before the field again becomes positive.

For $L=1024$ the picture is quite different. In the snapshot we see 
a large number of growing clusters that have nucleated at different times 
during the negative-field time interval. Some of these have already coalesced 
by the time the snapshot was captured, 
while others are still growing independently.    
From the time series it is seen that this multi-droplet 
switching mode leads to a nearly deterministic evolution of the total 
magnetization, with 
the underlying stochasticity only evident in the slight variations of the 
minimum magnetization values from period to period.  
This switching process is well described by the 
Kolmogorov-Johnson-Mehl-Avrami (KJMA) 
approximation \cite{RIKV94A,BIND16,KOLM37,JOHN39,AVRAMI,RAMO99}. 

Magnetization reversal from the favored to the disfavored direction is only possible while 
the total applied field, 
$H(t) + H_b$, has the opposite sign of the bias, $H_b$. This implies that 
$-1 < H_b/H(t) \le 0$. Switching from the favored phase to the disfavored one on average takes 
longer time than switching in the opposite direction. 
Thus, the time the system can spend in the disfavored phase 
during each period must be less than or equal to the time that the field has the 
disfavored direction, 
\begin{equation}
t_{\rm Dmax} 
= \frac{P}{2} \left[ 1 - \frac{2}{\pi} \sin^{-1} \left( \frac{|H_b|}{H_0} \right) \right] \;. 
\label{eq:tnmax}
\end{equation}
In this limit of long period and weak bias, $\langle Q \rangle$ is simply determined by 
the sign of $H_b$ and the difference between the fractions of the period that the total field 
has the same and the opposite sign as $H_b$, respectively. This yields  
\begin{equation}
\langle Q \rangle  
\approx 
\frac{2 m_{0}}{\pi} \sin^{-1} \left( \frac{H_b}{H_0} \right) \;,
\label{eq:QQ}
\end{equation}
which is symmetric under simultaneous reversal of $H_b$ and $\langle Q \rangle$.
Here, $m_{0}$ is the magnitude of the magnetization in the favored phase. 
This approximation represents a lower bound on the magnitudes of $\langle Q \rangle$ 
and $\chi^b$ \cite{RIEG17B}. The former 
is included as a dashed curve in Fig.~\protect\ref{fig:128vsh}(a). 
However, the bounds depend on the waveform of the oscillating field, and as we 
show in Appendix \ref{sec:SQW}, they vanish in the case of a square-wave field. 

The corrections to this approximation are of $O \left(t_{\rm FD}(H_b,H_0)/P \right)$, where 
$t_{\rm FD}(H_b,H_0)$ is the average time it takes the magnetization to switch to the 
disfavored direction, {\it after} the total applied field has changed sign. For $|H_b| \ll H_0$, the 
correction vanishes as $1/P$, as seen in Fig.\ \ref{fig:128vsh}(a). 
However, for larger $|H_b|$, $t_{\rm FD}(H_b,H_0) \sim P$, and 
the ``correction" becomes the dominant part of $\langle Q \rangle$, determining the sideband 
peak positions, $H_b^{\rm peak}$. 
The details are discussed below in Sec.\ \ref{sec:HP}.

\subsection{Dependence on $H_b$, $P$, and $L$}
\label{sec:HP}

\begin{figure}[ht]
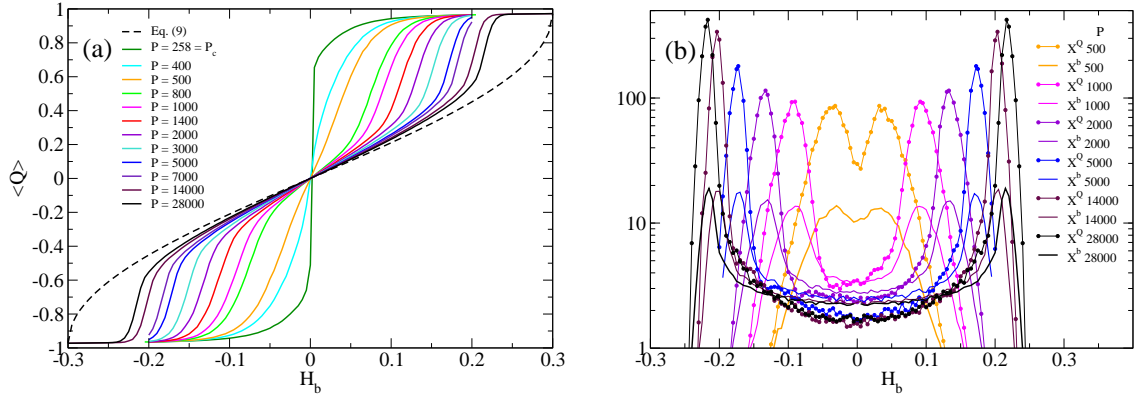

\begin{center}
\vspace*{0.8truecm}
\includegraphics[angle=0,width=.45\textwidth]{Q_LongP_L128X.eps} 
\hspace*{0.4truecm}
\includegraphics[angle=0,width=.42\textwidth]{Peaks_LongP_L128_PAPX.eps} 
%\includegraphics[angle=0,width=.42\textwidth]{Peaks_LongP_L128.eps} 
%\vspace*{3.0truecm}
%\includegraphics[angle=0,width=.45\textwidth]{deriv128P.eps} 
\end{center}
\vspace*{-0.3truecm}
\caption[]{
%\baselineskip=0.3truecm
\baselineskip=0.15truecm
Results for $L=128$ and a range of periods between $P_c = 258$ and 
$P = 28,000$. 
(a)
The order parameter $\langle Q \rangle$ vs $H_{b}$. Error bars 
are on the order of the line thickness. 
The dashed curve is the weak-bias, long-period approximation of Eq.\ (\protect\ref{eq:QQ}). 
(b) 
The scaled 
variance $\chi_{L}^{Q}$ and the susceptibility $\chi_{L}^{b}$ vs  
$H_{b}$. The 
sideband peaks occur at values of $H_{b}$ that increase with $P$. 
For clarity, data for some values of $P$ are omitted in (b), 
including a narrow critical peak for 
$P=P_c=258$ at $H_{b}=0$ and a broad central peak for $P=400$. 
}
\label{fig:128vsh}
\end{figure}

Results for $L=128$ and a range of periods between $P_c = 258$ and $P = 28,000$  
are shown in Fig.\ \ref{fig:128vsh}. In the critical region, $H_b$ is the field conjugate to 
$\langle Q \rangle$ \cite{ROBB07,GALL12,IDIG12,ROBB14,BERG13}.
At $P = P_c$, $\langle Q \rangle$ therefore vanishes in a singular fashion 
as $H_b$ approaches zero. On the scale of Fig.\ \ref{fig:128vsh}(a), 
this singularity appears as a jump in $\langle Q \rangle$ at 
$H_b = 0$ for $P=P_c$, resulting in very narrow central peaks 
in both $\chi_L^b$ and $\chi_L^Q$. 
We also found broad central peaks in both quantities for $P = 400$, which are due to 
finite-size broadening of the critical region for this relatively modest system size. 
For clarity, these central peaks are not included in Fig.\ \ref{fig:128vsh}(b). 
Beyond $P = 500$, $\langle Q \rangle$ becomes linear for small $H_b$, with a slope that 
approaches that of the asymptotic approximation in Eq.\ (\ref{eq:QQ}) as $P$ increases. 
Simultaneously, the peaks in $\chi_L^b$ and $\chi_L^Q$ increase in height, and their positions 
$H_b^{\rm peak}$ move in the directions of $\pm H_0$, as seen in Fig.\ \ref{fig:128vsh}(b). 
[For clarity, some of the values of $P$ included in 
Fig.\ \ref{fig:128vsh}(a) are excluded from Fig.\ \ref{fig:128vsh}(b).]

\begin{figure}[ht]
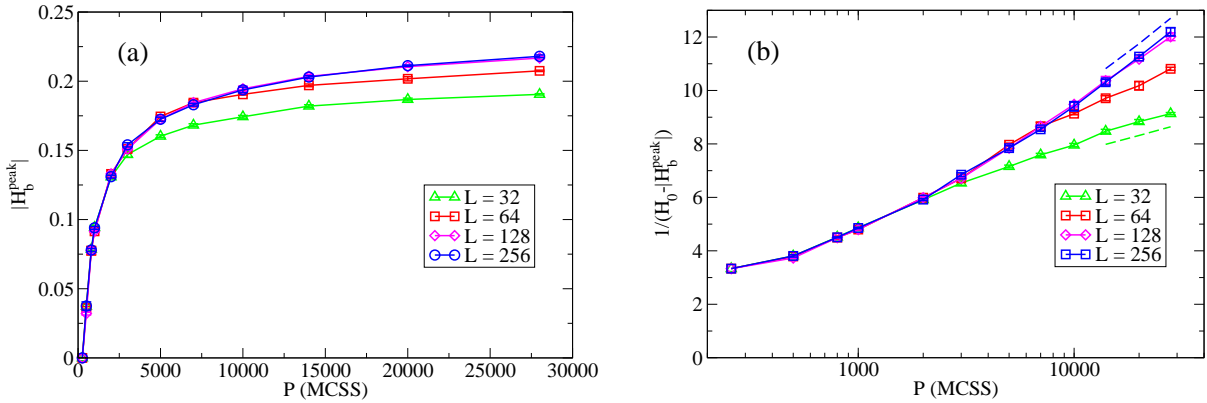

\begin{center}
\vspace*{0.8truecm}
\includegraphics[angle=0,width=.48\textwidth]{PeakPos_vs_PX.eps} 
\hspace*{0.4truecm}
\includegraphics[angle=0,width=.45\textwidth]{PeakPos_vs_logPX.eps} 
\end{center}
\vspace*{-0.3truecm}
\caption[]{
%\baselineskip=0.3truecm
\baselineskip=0.15truecm
Peak positions $|H_b^{\rm peak}|$ as defined by the maxima of the scaled 
variance $\chi_{L}^{Q}$, shown vs period length $P \ge P_c$. 
(a)
$|H_b^{\rm peak}|$ vs $P$, plotted on linear scales. 
(b)
The peak positions plotted as $1/(H_0 - |H_b^{\rm peak}|)$ vs $\log P$, 
as suggested by Eq.~(\protect\ref{eq:pekreq1}). The blue and green dashed lines represent the 
slopes of the curves between $P=14,000$ and 28,000 for $L=256$ and $L=32$, 
respectively. The ratio of the slopes is approximately 2.867, close to the 3/1 ratio expected 
from droplet theory. 
This figure is analogous to Fig.\ 2 of Ref.\ \protect\cite{RIKV94A}. 
}
\label{fig:PeakPos}
\end{figure}

The magnitudes of the peak positions, $|H_b^{\rm peak}|$, are plotted vs $P$ for different 
values of $L$ in Fig.\ \ref{fig:PeakPos}(a). We note two main features. First, 
$|H_b^{\rm peak}|$ increases quite rapidly with $P$ for relatively short periods, and 
much more slowly for longer periods. 
This behavior is consistent with the experimental data shown in Fig.\ 2 of Ref.\ \cite{RIEG17}. 
Second, finite-size effects are essentially 
negligible for $P \lesssim 2000$, as already shown in 
Fig.\ \ref{fig:size1} for $P=1000$. For longer periods, $|H_b^{\rm peak}|$ 
increases with $L$ for smaller sizes, and then becomes size independent for larger $L$. 

\begin{figure}[ht]
\begin{center}
\vspace*{0.8truecm}
\includegraphics[angle=0,width=.7\textwidth]{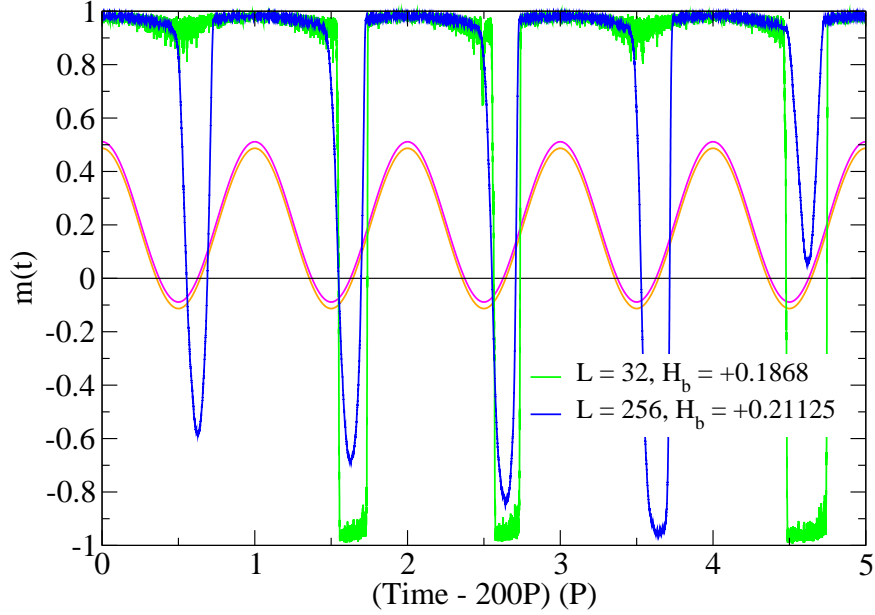} 
%\hspace*{0.4truecm}
%\includegraphics[angle=0,width=.45\textwidth]{PeakPos_vs_logP.eps} 
\end{center}
\vspace*{-0.3truecm}
\caption[]{
%\baselineskip=0.3truecm
\baselineskip=0.15truecm
Time series of $m(t)$ over five periods with $P = 20,000$, following a 200$P$ 
stabilization run. Data are shown at their respective values of $H_b^{\rm peak}$ for 
$L = 32$ (green) and 256 (blue). The corresponding values of the total applied field, 
$H(t) + H_b^{\rm peak}$ are also shown in orange and magenta, respectively. 
The wave forms of $m(t)$, characteristic of single-droplet and multidroplet switching are seen 
for $L=32$ and 256, respectively.   
}
\label{fig:P20000}
\end{figure}

In order to explain this behavior quantitatively, we first recall from the time series 
shown in Fig.\ \ref{fig:tenP} that for bias near $|H_b^{\rm peak}|$, the time it takes 
$m(t)$ to change significantly toward the disfavored sign is on the order of a finite fraction of $P$. 
For stronger bias, the total field driving the magnetization toward the disfavored sign
is too weak and consequently the time required for switching is much longer than $P$, 
so that reliable 
magnetization reversal does not occur. For weaker bias, the field in the disfavored direction is relatively strong, and 
complete and reliable magnetization reversal takes place on a timescale significantly shorter than 
$P$.  In other words, the peak positions correspond to bias values that produce 
magnetization reversal on a timescale of $P$. Equations 
for magnetization switching rates by the stochastic single-particle mechanism that 
dominates for small systems 
[Eq.\ (\ref{eq:IH})] and the nearly deterministic multidroplet mechanism 
that dominates for large systems [Eq.\ (\ref{eq:KJMA})] 
are found in Appendix \ref{sec:mech}.
The nucleation rate for droplets of the disfavored phase 
varies very strongly with the oscillating field, having appreciable values 
only in a narrow window near the maximum field in the disfavored direction, 
$|H| = H_0 - |H_b^{\rm peak}|$. 
Using this value of $|H|$ and ignoring less important prefactors, we can 
use these equations to write the following requirement for $|H_b^{\rm peak}|$: 
\begin{equation}
L^{-a} \exp \left( \frac{1}{b} \frac{\Xi_0}{H_0 - |H_b^{\rm peak}| } \right) \sim P \;,
\label{eq:pekreq1}
\end{equation}
with $a=2$ and $b=1$ for single-droplet switching, and $a=0$ and $b=3$ for 
multidroplet switching. 
The meaning of the constant $\Xi_0 \approx 0.506$ is explained in Appendix \ref{sec:mech}. 
In either case, this equation is equivalent to a statement that $|H_b^{\rm peak}|$ 
should approach $H_0$ asymptotically as $1/ \log P$ for long periods. 
(A caveat to this statement for the case of extremely long periods is discussed in 
Appendix \ref{sec:CAV}.) 
Plotting $1/(H_0 - |H_b^{\rm peak}|)$ 
vs $\log P$ therefore should produce straight lines for large values of $P$. 
The ratio between the slopes of the lines representing multidroplet switching for large $L$ 
and those representing single-droplet switching for small $L$ should be 3/1. 
Such a plot is presented in Fig.\ \ref{fig:PeakPos}(b). The slope ratio between the curves 
representing $L=256$ and $L=32$ in the long-$P$ regime is approximately 2.867, 
consistent with the theoretical prediction. This conclusion is confirmed by the short 
time series of $m(t)$ for $P=20,000$ for these two system sizes, shown in 
Fig.\ \ref{fig:P20000}. In the switching regions, the smaller system displays the 
stochastic, square wave form characteristic of single-droplet switching \cite{SIDE98A}, 
while the larger system shows the continuous 
wave form characteristic of multidroplet switching \cite{SIDE99}. 

To further support our conclusions, we calculated the transition times and the order parameter 
in the multidroplet regime for the mathematically simpler case, 
in which the sinusoidally oscillating field has 
been replaced by a square-wave field. The details of the calculations are given in Appendix 
\ref{sec:SQW}. In Fig.\ \ref{fig:Qsq} we show that there is very good agreement between the theoretically 
calculated $\langle Q \rangle$ and the simulations, particularly when 
$|H_b| \lesssim |H_b ^{\rm peak}|$.

\section{Summary and Conclusion}
\label{sec:CONC}

Riego {\it et al.} \cite{RIEG17} recently presented experimental data on Co films with a 
single, in-plane magnetic easy axis, which were subjected to a slowly oscillating magnetic 
field with an added constant bias. In this paper we have presented kinetic MC simulations and 
theoretical analysis of a two-dimensional Ising ferromagnet with only 
nearest-neighbor interactions, designed to closely mimic the experimental setup. 
At zero bias, such systems exhibit a dynamic phase transition (DPT) at a critical period $P_c$, 
where the period-averaged magnetization $\langle Q \rangle$ vanishes in a singular fashion. 
It has previously been shown that the DPT belongs to the equilibrium Ising universality class, 
with $P$ playing the role of temperature and the bias $H_b$ being the field conjugate to 
$\langle Q \rangle$. 
Following Riego {\it et al.} \cite{RIEG17}, 
we studied the dynamics of the system at values of $P$ {\it above} 
$P_c$, and in agreement with the experiments we found that $\langle Q \rangle$ exhibits a strong bias dependence and fluctuation peaks at nonzero values of 
$H_b$, symmetrically located around zero bias. 

Since the simulated system has only nearest-neighbor interactions, our results show that the 
experimental results are {\it not} due to any residual magnetostatic interactions. 
The simulational approach also enables studies of the effects of finite system size. 
We found that, at fixed $P$, finite-size effects saturate beyond a $P$-dependent size limit. 
Using the droplet theory of magnetization switching, we conclude that this saturation occurs 
at the crossover between two different dynamic regimes. For small systems, the magnetization 
switching from the favored to the disfavored direction occurs by a 
stochastic single-droplet mechanism. For large systems, the switching occurs by the size-independent 
and nearly deterministic KJMA 
mechanism, which involves a large 
number of simultaneously nucleating and growing droplets.  
We therefore conclude that this ``sideband" phenomenon for 
supercritical values of $P$ is {\it not} a critical phenomenon, but rather a stochastic-resonance 
phenomenon. 
We believe these insights will be important for the design and analysis of devices that involve 
magnetization reversal by time-varying fields, such as memory elements, switches,  
and actuators.

\section*{Acknowledgments}
We thank A.\ Berger for providing data from Ref.\ \cite{RIEG17} before publication, 
and for useful comments on an earlier version of this paper. 
G.M.B.\ is grateful for the hospitality of the Physics Department at Florida State University, 
where her stay was supported in part by the American Physical Society International 
Research Travel Award Program (IRTAP).  
P.A.R.\ acknowledges partial support by U.S. National Science Foundation Grant No. DMR-1104829. 

\appendix

\section{Mechanisms of magnetization reversal} 
\label{sec:mech}
When a $d$-dimensional Ising ferromagnet below its critical temperature is subjected to 
the reversal of an 
applied field of magnitude $|H|$, the homogeneous nucleation rate per unit system volume 
for droplets of the new equilibrium phase is given 
by \cite{SIDE99,SIDE98A,RIKV94A,LANG67,LANG69,GNW80,BIND16} 
\begin{equation}
I(H) \approx B(T) |H|^K \exp \left[ - \frac{\Xi_0 (T)}{|H|^{d-1}} \right] \;,
\label{eq:IH}
\end{equation}
where $B(T)$ is a non-universal function of $T$. For $d=2$, $K=3$, and  
$\Xi_0(0.8 T_c) \approx 0.506$ (which includes a factor of $1/T$)  \cite{SIDE99}. 
The argument of the exponential function is the negative of the 
free energy of a critical droplet of the equilibrium phase, divided by $T$. 
The inverse of $L^d I(H)$ is 
the average time between random nucleation events for a system of size $L$. 

{\bf Single-droplet reversal mechanism:}
Under conditions of small system and/or moderately weak field, 
the time it takes for the first nucleated droplet to grow to fill the system 
is much shorter than the average nucleation time. As a result, the magnetization reversal is 
completed by this single, first droplet. 

{\bf Multidroplet reversal mechanism:}
Under conditions of large system and/or moderately strong field, 
the average time between nucleation events is less than the time it would take the first 
nucleated droplet to grow to fill the system. Therefore, many droplets nucleate and grow 
independently in different parts of the system until they coalesce and collectively fill the system. 
The result is a gradual and nearly deterministic growth of the new phase through a multidroplet 
process, well described by the KJMA 
approximation \cite{RIKV94A,BIND16,KOLM37,JOHN39,AVRAMI,RAMO99}. 
The characteristic reversal time is independent of the system size and given by 
\begin{equation}
\langle \tau(H) \rangle \propto \left[ v^d I(H) \right] ^{-1/(d+1)} \;,
\label{eq:KJMA}
\end{equation} 
where the propagation velocity of  the droplet surface, $v$, 
is proportional to $|H|$ in this parameter range 
\cite{RIKV00B} as expected from the Lifshitz-Allen-Cahn approximation 
\cite{GUNT83A,LIFS62,ALLE79}. 

\section{Extremely long periods} 
\label{sec:CAV}

If the radius of the critical droplet reaches a size of about $L/2$, it will not fit in the $L \times L$ 
system, and a new regime, called the {\it coexistence regime\/}, is entered \cite{RIKV94A}. In this 
regime, the droplet is replaced by a slab of the equilibrium phase, and the nucleation time no longer 
depends on $|H|$, but increases exponentially with $L^{d-1}$. 
The critical droplet radius in $d$ dimensions is given by  
\cite{RIKV94A},
\begin{equation}
R_c \approx 
\left( \frac{(d-1) T \Xi_0}{2 m_0 \Omega_d}  \right)^{1/d} \frac{1}{|H|} \;,
\label{eq:Rc}
\end{equation}
where $\Omega_d$ is the volume of the critical droplet, divided by $R_c^d$. 
Numerical  values for the constants with $d=2$ at 
$T=0.8T_c \approx 1.815$ are found in Table I of Ref.\ \cite{SIDE99}: 
$\Xi_0 \approx 0.506$ and $\Omega_2 \approx 3.152$. 
(The factor $T$ is included in the numerator to cancel the factor $1/T$ in $\Xi_0$.)
Thus we have 
\begin{equation}
R_c \approx \frac{0.388}{|H|} \approx \frac{L}{2} \;.
\label{eq:RcL}
\end{equation}
Replacing $|H|$ by $H_0 - |H_b|$ and setting $L=32$, we thus find 
$1/(H_0 - |H_b|) \approx 41.3$. Finally, linearly extrapolating the large-$P$ data for $L=32$ in 
Fig.\ \ref{fig:PeakPos}(b), we find that the single-droplet result from Eq.\ (\ref{eq:pekreq1}) 
should remain valid for periods up to approximately $10^{19 \pm 2}$. 
[The uncertainty in the exponent is the result of assuming a $10\%$ uncertainty in the estimate of 
$1/(H_0 - |H_b|)$.]
Beyond this limit, $H_0 - |H_b|$ should remain independent of $P$, at a value of $O(1/L)$. 
For larger $L$, the single-droplet result should be valid up to even longer periods. 
We do not expect that these extremely long periods should be of great experimental relevance for 
macroscopic systems. However, for nanoscopic systems the coexistence regime may be 
observable with experimentally accessible periods.

\section{Square-wave oscillating field} 
\label{sec:SQW}
Now, instead of a sinusoidally oscillating field, consider a square-wave field, such that 
$H(t) = +H_0$ during one half-period, and $-H_0$ during the other. 
Since the times that the total field is parallel and antiparallel to $H_b$ now each equal $P/2$, 
the equivalent of the long-period, weak-bias approximation of Eq.\ (\ref{eq:QQ}) becomes 
$\langle Q \rangle \approx 0$. Therefore, the value of $\langle Q \rangle$ for finite $P$ and 
weak $H_b$ is determined by the difference between the average magnetization reversal times 
following a change of the total field from the favored to the disfavored direction, and the 
opposite. Since the total field now has its full favored or disfavored strength during the whole 
half-period, these average switching times will be shorter than the corresponding times in the 
sinusoidally oscillating field case. With a square-wave field 
of amplitude $H_0 = 0.3$ at $0.8 T_c$ under Glauber dynamics, the critical period has 
been measured by MC simulations as $P_c \approx 137$ \cite{KORN00}. 
To calculate the transition times for a two-dimensional system in the multidroplet regime, 
we will again assume $H_b \ge 0$ for concreteness. 

From Eqs.\ (\ref{eq:IH}) and (\ref{eq:KJMA}) with $|H| = H_0 - H_b$, 
we obtain the characteristic timescale for 
transitions from the favored (parallel to the bias field) to the disfavored 
magnetization direction, after the total applied field has changed sign as 
\begin{equation}
t_{\rm FD}(H_b,H_0) = \tau_0 \left( \frac{1}{1 - H_b/H_0} \right)^{5/3} 
\exp \left(\frac{\Xi_0}{3 H_0} \frac{H_b/H_0}{1 - H_b/H_0} \right)  \ge \tau_0 \;,
\label{eq:tfd}
\end{equation}
where $\tau_0$ is the magnetization reversal time for $H_b = 0$.  
Analogously, the switching time from the disfavored to the favored magnetization direction is 
\begin{equation}
t_{\rm DF}(H_b,H_0) = \tau_0 \left( \frac{1}{1 + H_b/H_0} \right)^{5/3} 
\exp \left( - \frac{\Xi_0}{3 H_0} \frac{H_b/H_0}{1 + H_b/H_0} \right)  \le \tau_0 \;.
\label{eq:tdf}
\end{equation}
Both $t_{\rm FD}$ and $t_{\rm DF}$ reduce to $\tau_0 \approx 74.6$ \cite{SIDE99} 
for $H_b = 0$. 

\begin{figure}[ht]
\begin{center}
\vspace*{0.8truecm}
\includegraphics[angle=0,width=.7\textwidth]{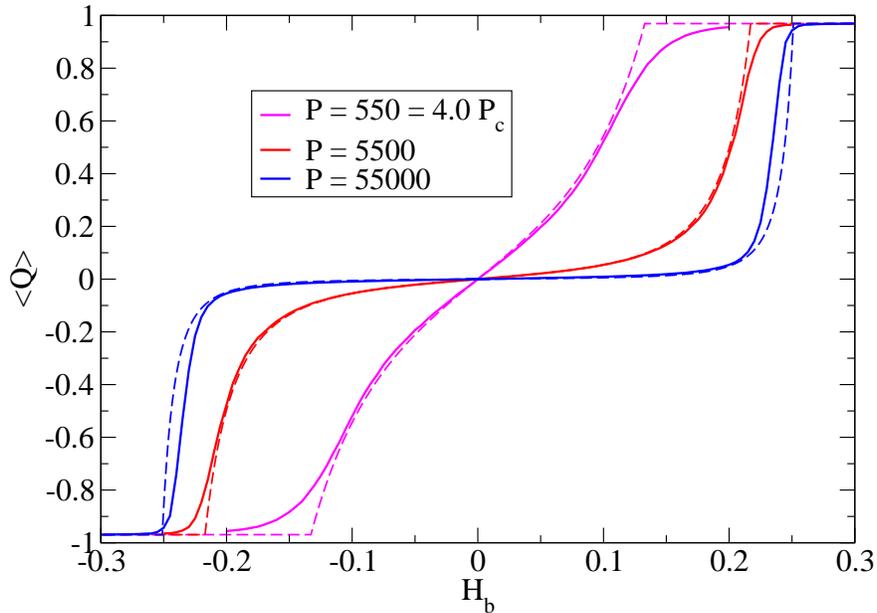} 
%\hspace*{0.4truecm}
%\includegraphics[angle=0,width=.45\textwidth]{PeakPos_vs_logP.eps} 
\end{center}
\vspace*{-0.3truecm}
\caption[]{
%\baselineskip=0.3truecm
\baselineskip=0.15truecm
Simulated results (solid) and approximate theoretical results from 
Eqs.\ (\ref{eq:tfd}) -- (\ref{eq:Qsq}) (dashed) for 
$\langle Q \rangle$ with a square-wave field of amplitude $H_0 = 0.3$. 
System size $L= 128$ and three different field periods $P$. 
In a square-wave field, $P_c \approx 137$ \protect\cite{KORN00}. 
}
\label{fig:Qsq}
\end{figure}

The order parameter $\langle Q \rangle$ is determined 
by $P$ and the difference between $t_{\rm FD}$ and $t_{\rm DF}$ as 
\begin{equation}
\langle Q \rangle
\approx
\left\{
\begin{array}{lll}
2 m_0 \frac{t_{\rm FD} - t_{\rm DF}}{P} 
& \mbox{for} & t_{\rm FD} \leq \frac{P}{2} \\
m_0  
& \mbox{for} & t_{\rm FD} > \frac{P}{2}
\end{array}
\right.
\label{eq:Qsq}
\end{equation}
This approximation is shown together with simulation results in Fig.\ \ref{fig:Qsq}. 
The agreement is very good for $|H_b| \lesssim |H_b^{\rm peak}|$. 

%\clearpage

%\bibliography{metastab}

\begin{thebibliography}{34}%
\makeatletter
\providecommand \@ifxundefined [1]{%
 \@ifx{#1\undefined}
}%
\providecommand \@ifnum [1]{%
 \ifnum #1\expandafter \@firstoftwo
 \else \expandafter \@secondoftwo
 \fi
}%
\providecommand \@ifx [1]{%
 \ifx #1\expandafter \@firstoftwo
 \else \expandafter \@secondoftwo
 \fi
}%
\providecommand \natexlab [1]{#1}%
\providecommand \enquote  [1]{``#1''}%
\providecommand \bibnamefont  [1]{#1}%
\providecommand \bibfnamefont [1]{#1}%
\providecommand \citenamefont [1]{#1}%
\providecommand \href@noop [0]{\@secondoftwo}%
\providecommand \href [0]{\begingroup \@sanitize@url \@href}%
\providecommand \@href[1]{\@@startlink{#1}\@@href}%
\providecommand \@@href[1]{\endgroup#1\@@endlink}%
\providecommand \@sanitize@url [0]{\catcode `\\12\catcode `\$12\catcode
  `\&12\catcode `\#12\catcode `\^12\catcode `\_12\catcode `\%12\relax}%
\providecommand \@@startlink[1]{}%
\providecommand \@@endlink[0]{}%
\providecommand \url  [0]{\begingroup\@sanitize@url \@url }%
\providecommand \@url [1]{\endgroup\@href {#1}{\urlprefix }}%
\providecommand \urlprefix  [0]{URL }%
\providecommand \Eprint [0]{\href }%
\providecommand \doibase [0]{http://dx.doi.org/}%
\providecommand \selectlanguage [0]{\@gobble}%
\providecommand \bibinfo  [0]{\@secondoftwo}%
\providecommand \bibfield  [0]{\@secondoftwo}%
\providecommand \translation [1]{[#1]}%
\providecommand \BibitemOpen [0]{}%
\providecommand \bibitemStop [0]{}%
\providecommand \bibitemNoStop [0]{.\EOS\space}%
\providecommand \EOS [0]{\spacefactor3000\relax}%
\providecommand \BibitemShut  [1]{\csname bibitem#1\endcsname}%
\let\auto@bib@innerbib\@empty
%</preamble>
\bibitem [{\citenamefont {Tom{\'e}}\ and\ \citenamefont
  {de~Oliveira}(1990)}]{TOME90}%
  \BibitemOpen
  \bibfield  {author} {\bibinfo {author} {\bibfnamefont {T.}~\bibnamefont
  {Tom{\'e}}}\ and\ \bibinfo {author} {\bibfnamefont {M.~J.}\ \bibnamefont
  {de~Oliveira}},\ }\bibfield  {title} {\enquote {\bibinfo {title} {Dynamic
  phase transition in the kinetic Ising model under a time-dependent
  oscillating field},}\ }\href@noop {} {\bibfield  {journal} {\bibinfo
  {journal} {Phys.\ Rev.\ A}\ }\textbf {\bibinfo {volume} {41}},\ \bibinfo
  {pages} {4251} (\bibinfo {year} {1990})}\BibitemShut {NoStop}%
\bibitem [{\citenamefont {Rao}\ \emph {et~al.}(1990)\citenamefont {Rao},
  \citenamefont {Krishnamurthy},\ and\ \citenamefont {Pandit}}]{RAO90B}%
  \BibitemOpen
  \bibfield  {author} {\bibinfo {author} {\bibfnamefont {M.}~\bibnamefont
  {Rao}}, \bibinfo {author} {\bibfnamefont {H.~R.}\ \bibnamefont
  {Krishnamurthy}}, \ and\ \bibinfo {author} {\bibfnamefont {R.}~\bibnamefont
  {Pandit}},\ }\bibfield  {title} {\enquote {\bibinfo {title} {Magnetic
  hysteresis in two model spin systems},}\ }\href@noop {} {\bibfield  {journal}
  {\bibinfo  {journal} {Phys.\ Rev.\ B}\ }\textbf {\bibinfo {volume} {42}},\
  \bibinfo {pages} {856} (\bibinfo {year} {1990})}\BibitemShut {NoStop}%
\bibitem [{\citenamefont {Lo}\ and\ \citenamefont {Pelcovits}(1990)}]{LO90}%
  \BibitemOpen
  \bibfield  {author} {\bibinfo {author} {\bibfnamefont {W.~S.}\ \bibnamefont
  {Lo}}\ and\ \bibinfo {author} {\bibfnamefont {R.~A.}\ \bibnamefont
  {Pelcovits}},\ }\bibfield  {title} {\enquote {\bibinfo {title} {Ising model
  in a time-dependent magnetic field},}\ }\href@noop {} {\bibfield  {journal}
  {\bibinfo  {journal} {Phys.\ Rev.\ A}\ }\textbf {\bibinfo {volume} {42}},\
  \bibinfo {pages} {7471} (\bibinfo {year} {1990})}\BibitemShut {NoStop}%
\bibitem [{\citenamefont {Chakrabarti}\ and\ \citenamefont
  {Acharyya}(1999)}]{CHAK99}%
  \BibitemOpen
  \bibfield  {author} {\bibinfo {author} {\bibfnamefont {B.}~\bibnamefont
  {Chakrabarti}}\ and\ \bibinfo {author} {\bibfnamefont {M.}~\bibnamefont
  {Acharyya}},\ }\bibfield  {title} {\enquote {\bibinfo {title} {Dynamic
  transitions and hysteresis},}\ }\href@noop {} {\bibfield  {journal} {\bibinfo
   {journal} {Rev.\ Mod.\ Phys.}\ }\textbf {\bibinfo {volume} {71}},\ \bibinfo
  {pages} {847} (\bibinfo {year} {1999})}\BibitemShut {NoStop}%
\bibitem [{\citenamefont {Sides}\ \emph
  {et~al.}(1998{\natexlab{a}})\citenamefont {Sides}, \citenamefont {Rikvold},\
  and\ \citenamefont {Novotny}}]{SIDE98}%
  \BibitemOpen
  \bibfield  {author} {\bibinfo {author} {\bibfnamefont {S.~W.}\ \bibnamefont
  {Sides}}, \bibinfo {author} {\bibfnamefont {P.~A.}\ \bibnamefont {Rikvold}},
  \ and\ \bibinfo {author} {\bibfnamefont {M.~A.}\ \bibnamefont {Novotny}},\
  }\bibfield  {title} {\enquote {\bibinfo {title} {Kinetic Ising model in an
  oscillating field: Finite-size scaling at the dynamic phase transition},}\
  }\href@noop {} {\bibfield  {journal} {\bibinfo  {journal} {Phys.\ Rev.\
  Lett.}\ }\textbf {\bibinfo {volume} {81}},\ \bibinfo {pages} {834}
  (\bibinfo {year} {1998}{\natexlab{a}})}\BibitemShut {NoStop}%
\bibitem [{\citenamefont {Sides}\ \emph {et~al.}(1999)\citenamefont {Sides},
  \citenamefont {Rikvold},\ and\ \citenamefont {Novotny}}]{SIDE99}%
  \BibitemOpen
  \bibfield  {author} {\bibinfo {author} {\bibfnamefont {S.~W.}\ \bibnamefont
  {Sides}}, \bibinfo {author} {\bibfnamefont {P.~A.}\ \bibnamefont {Rikvold}},
  \ and\ \bibinfo {author} {\bibfnamefont {M.~A.}\ \bibnamefont {Novotny}},\
  }\bibfield  {title} {\enquote {\bibinfo {title} {Kinetic Ising model in an
  oscillating field: Avrami theory for the hysteretic response and finite-size
  scaling for the dynamic phase transition},}\ }\href@noop {} {\bibfield
  {journal} {\bibinfo  {journal} {Phys.\ Rev.\ E}\ }\textbf {\bibinfo {volume}
  {59}},\ \bibinfo {pages} {2710} (\bibinfo {year} {1999})}\BibitemShut
  {NoStop}%
\bibitem [{\citenamefont {Korniss}\ \emph {et~al.}(2000)\citenamefont
  {Korniss}, \citenamefont {White}, \citenamefont {Rikvold},\ and\
  \citenamefont {Novotny}}]{KORN00}%
  \BibitemOpen
  \bibfield  {author} {\bibinfo {author} {\bibfnamefont {G.}~\bibnamefont
  {Korniss}}, \bibinfo {author} {\bibfnamefont {C.~J.}\ \bibnamefont {White}},
  \bibinfo {author} {\bibfnamefont {P.~A.}\ \bibnamefont {Rikvold}}, \ and\
  \bibinfo {author} {\bibfnamefont {M.~A.}\ \bibnamefont {Novotny}},\
  }\bibfield  {title} {\enquote {\bibinfo {title} {Dynamic phase transition,
  universality, and finite-size scaling in the two-dimensional kinetic Ising
  model in an oscillating field},}\ }\href@noop {} {\bibfield  {journal}
  {\bibinfo  {journal} {Phys.\ Rev.\ E}\ }\textbf {\bibinfo {volume} {63}},\
  \bibinfo {pages} {016120} (\bibinfo {year} {2000})}\BibitemShut
  {NoStop}%
\bibitem [{\citenamefont {Robb}\ \emph {et~al.}(2007)\citenamefont {Robb},
  \citenamefont {Rikvold}, \citenamefont {Berger},\ and\ \citenamefont
  {Novotny}}]{ROBB07}%
  \BibitemOpen
  \bibfield  {author} {\bibinfo {author} {\bibfnamefont {D.~T.}\ \bibnamefont
  {Robb}}, \bibinfo {author} {\bibfnamefont {P.~A.}\ \bibnamefont {Rikvold}},
  \bibinfo {author} {\bibfnamefont {A.}~\bibnamefont {Berger}}, \ and\ \bibinfo
  {author} {\bibfnamefont {M.~A.}\ \bibnamefont {Novotny}},\ }\bibfield
  {title} {\enquote {\bibinfo {title} {Conjugate field and
  fluctuation-dissipation relation for the dynamic phase transition in the
  two-dimensional kinetic Ising model},}\ }\href@noop {} {\bibfield  {journal}
  {\bibinfo  {journal} {Phys.\ Rev.\ E}\ }\textbf {\bibinfo {volume} {76}},\
  \bibinfo {pages} {021124} (\bibinfo {year} {2007})}\BibitemShut {NoStop}%
\bibitem [{\citenamefont {Buend\'{\i}a}\ and\ \citenamefont
  {Rikvold}(2008)}]{BUEN08}%
  \BibitemOpen
  \bibfield  {author} {\bibinfo {author} {\bibfnamefont {G.~M.}\ \bibnamefont
  {Buend\'{\i}a}}\ and\ \bibinfo {author} {\bibfnamefont {P.~A.}\ \bibnamefont
  {Rikvold}},\ }\bibfield  {title} {\enquote {\bibinfo {title} {Dynamic phase
  transition in the two-dimensional kinetic Ising model in an oscillating
  field: Universality with respect to the stochastic dynamics},}\ }\href
  {\doibase 10.1103/PhysRevE.78.051108} {\bibfield  {journal} {\bibinfo
  {journal} {Phys. Rev. E}\ }\textbf {\bibinfo {volume} {78}},\ \bibinfo
  {pages} {051108} (\bibinfo {year} {2008})}\BibitemShut {NoStop}%
\bibitem [{\citenamefont {Park}\ and\ \citenamefont
  {Pleimling}(2013)}]{PARK13}%
  \BibitemOpen
  \bibfield  {author} {\bibinfo {author} {\bibfnamefont {H.}~\bibnamefont
  {Park}}\ and\ \bibinfo {author} {\bibfnamefont {M.}~\bibnamefont
  {Pleimling}},\ }\bibfield  {title} {\enquote {\bibinfo {title} {Dynamic phase
  transition in the three-dimensional kinetic Ising model in an oscillating
  field},}\ }\href {\doibase 10.1103/PhysRevE.87.032145} {\bibfield  {journal}
  {\bibinfo  {journal} {Phys. Rev. E}\ }\textbf {\bibinfo {volume} {87}},\
  \bibinfo {pages} {032145} (\bibinfo {year} {2013})}\BibitemShut {NoStop}%
\bibitem [{\citenamefont {Fujisaka}\ \emph {et~al.}(2001)\citenamefont
  {Fujisaka}, \citenamefont {Tutu},\ and\ \citenamefont {Rikvold}}]{FUJI01}%
  \BibitemOpen
  \bibfield  {author} {\bibinfo {author} {\bibfnamefont {H.}~\bibnamefont
  {Fujisaka}}, \bibinfo {author} {\bibfnamefont {H.}~\bibnamefont {Tutu}}, \
  and\ \bibinfo {author} {\bibfnamefont {P.~A.}\ \bibnamefont {Rikvold}},\
  }\bibfield  {title} {\enquote {\bibinfo {title} {Dynamic phase transition in
  a time-dependent Ginzburg-Landau model in an oscillating field},}\
  }\href@noop {} {\bibfield  {journal} {\bibinfo  {journal} {Phys.\ Rev.\ E}\
  }\textbf {\bibinfo {volume} {63}},\ \bibinfo {pages} {036109}
  (\bibinfo {year} {2001})};\ \bibinfo {note} {erratum: {\bf 63}, 059903 
(2001).}\BibitemShut {Stop}%
\bibitem [{\citenamefont {Gallardo}\ \emph {et~al.}(2012)\citenamefont
  {Gallardo}, \citenamefont {Idigoras}, \citenamefont {Landeros},\ and\
  \citenamefont {Berger}}]{GALL12}%
  \BibitemOpen
  \bibfield  {author} {\bibinfo {author} {\bibfnamefont {R.~A.}\ \bibnamefont
  {Gallardo}}, \bibinfo {author} {\bibfnamefont {O.}~\bibnamefont {Idigoras}},
  \bibinfo {author} {\bibfnamefont {P.}~\bibnamefont {Landeros}}, \ and\
  \bibinfo {author} {\bibfnamefont {A.}~\bibnamefont {Berger}},\ }\bibfield
  {title} {\enquote {\bibinfo {title} {Analytical derivation of critical
  exponents of the dynamic phase transition in the mean-field approximation},}\
  }\href {\doibase 10.1103/PhysRevE.86.051101} {\bibfield  {journal} {\bibinfo
  {journal} {Phys. Rev. E}\ }\textbf {\bibinfo {volume} {86}},\ \bibinfo
  {pages} {051101} (\bibinfo {year} {2012})}\BibitemShut {NoStop}%
\bibitem [{\citenamefont {Idigoras}\ \emph {et~al.}(2012)\citenamefont
  {Idigoras}, \citenamefont {Vavassori},\ and\ \citenamefont
  {Berger}}]{IDIG12}%
  \BibitemOpen
  \bibfield  {author} {\bibinfo {author} {\bibfnamefont {O.}~\bibnamefont
  {Idigoras}}, \bibinfo {author} {\bibfnamefont {P.}~\bibnamefont {Vavassori}},
  \ and\ \bibinfo {author} {\bibfnamefont {A.}~\bibnamefont {Berger}},\
  }\bibfield  {title} {\enquote {\bibinfo {title} {Mean field theory of dynamic
  phase transitions in ferromagnets},}\ }\href {\doibase
  http://dx.doi.org/10.1016/j.physb.2011.06.029} {\bibfield  {journal}
  {\bibinfo  {journal} {Physica B: Condensed Matter}\ }\textbf {\bibinfo
  {volume} {407}},\ \bibinfo {pages} {1377} (\bibinfo {year} {2012})}
%  \bibinfo {note} {8th International Symposium on Hysteresis Modeling and
%  Micromagnetics (HMM 2011)}
\BibitemShut {NoStop}%
\bibitem [{\citenamefont {Robb}\ and\ \citenamefont
  {Ostrander}(2014)}]{ROBB14}%
  \BibitemOpen
  \bibfield  {author} {\bibinfo {author} {\bibfnamefont {D.~T.}\ \bibnamefont
  {Robb}}\ and\ \bibinfo {author} {\bibfnamefont {A.}~\bibnamefont
  {Ostrander}},\ }\bibfield  {title} {\enquote {\bibinfo {title} {Extended
  order parameter and conjugate field for the dynamic phase transition in a
  Ginzburg-Landau mean-field model in an oscillating field},}\ }\href {\doibase
  10.1103/PhysRevE.89.022114} {\bibfield  {journal} {\bibinfo  {journal} {Phys.
  Rev. E}\ }\textbf {\bibinfo {volume} {89}},\ \bibinfo {pages} {022114}
  (\bibinfo {year} {2014})}\BibitemShut {NoStop}%
\bibitem [{\citenamefont {Robb}\ \emph {et~al.}(2008)\citenamefont {Robb},
  \citenamefont {Xu}, \citenamefont {Hellwig}, \citenamefont {McCord},
  \citenamefont {Berger}, \citenamefont {Novotny},\ and\ \citenamefont
  {Rikvold}}]{ROBB08}%
  \BibitemOpen
  \bibfield  {author} {\bibinfo {author} {\bibfnamefont {D.~T.}\ \bibnamefont
  {Robb}}, \bibinfo {author} {\bibfnamefont {Y.~H.}\ \bibnamefont {Xu}},
  \bibinfo {author} {\bibfnamefont {O.}~\bibnamefont {Hellwig}}, \bibinfo
  {author} {\bibfnamefont {J.}~\bibnamefont {McCord}}, \bibinfo {author}
  {\bibfnamefont {A.}~\bibnamefont {Berger}}, \bibinfo {author} {\bibfnamefont
  {M.~A.}\ \bibnamefont {Novotny}}, \ and\ \bibinfo {author} {\bibfnamefont
  {P.~A.}\ \bibnamefont {Rikvold}},\ }\bibfield  {title} {\enquote {\bibinfo
  {title} {Evidence for a dynamic phase transition in [Co/Pt]$_3$ magnetic
  multilayers},}\ }\href@noop {} {\bibfield  {journal} {\bibinfo  {journal}
  {Phys.\ Rev.\ B}\ }\textbf {\bibinfo {volume} {78}},\ \bibinfo {pages}
  {134422} (\bibinfo {year} {2008})}\BibitemShut {NoStop}%
\bibitem [{\citenamefont {Berger}\ \emph {et~al.}(2013)\citenamefont {Berger},
  \citenamefont {Idigoras},\ and\ \citenamefont {Vavassori}}]{BERG13}%
  \BibitemOpen
  \bibfield  {author} {\bibinfo {author} {\bibfnamefont {A.}~\bibnamefont
  {Berger}}, \bibinfo {author} {\bibfnamefont {O.}~\bibnamefont {Idigoras}}, \
  and\ \bibinfo {author} {\bibfnamefont {P.}~\bibnamefont {Vavassori}},\
  }\bibfield  {title} {\enquote {\bibinfo {title} {Transient behavior of the
  dynamically ordered phase in uniaxial cobalt films},}\ }\href {\doibase
  10.1103/PhysRevLett.111.190602} {\bibfield  {journal} {\bibinfo  {journal}
  {Phys. Rev. Lett.}\ }\textbf {\bibinfo {volume} {111}},\ \bibinfo {pages}
  {190602} (\bibinfo {year} {2013})}\BibitemShut {NoStop}%
\bibitem [{\citenamefont {Riego}\ \emph {et~al.}(2017)\citenamefont {Riego},
  \citenamefont {Vavassori},\ and\ \citenamefont {Berger}}]{RIEG17}%
  \BibitemOpen
  \bibfield  {author} {\bibinfo {author} {\bibfnamefont {P.}~\bibnamefont
  {Riego}}, \bibinfo {author} {\bibfnamefont {P.}~\bibnamefont {Vavassori}}, \
  and\ \bibinfo {author} {\bibfnamefont {A.}~\bibnamefont {Berger}},\
  }\bibfield  {title} {\enquote {\bibinfo {title} {Metamagnetic anomalies near
  dynamic phase transitions},}\ }\href {\doibase
  10.1103/PhysRevLett.118.117202} {\bibfield  {journal} {\bibinfo  {journal}
  {Phys. Rev. Lett.}\ }\textbf {\bibinfo {volume} {118}},\ \bibinfo {pages}
  {117202} (\bibinfo {year} {2017})}\BibitemShut {NoStop}%
\bibitem [{\citenamefont {Rikvold}\ \emph {et~al.}(1994)\citenamefont
  {Rikvold}, \citenamefont {Tomita}, \citenamefont {Miyashita},\ and\
  \citenamefont {Sides}}]{RIKV94A}%
  \BibitemOpen
  \bibfield  {author} {\bibinfo {author} {\bibfnamefont {P.~A.}\ \bibnamefont
  {Rikvold}}, \bibinfo {author} {\bibfnamefont {H.}~\bibnamefont {Tomita}},
  \bibinfo {author} {\bibfnamefont {S.}~\bibnamefont {Miyashita}}, \ and\
  \bibinfo {author} {\bibfnamefont {S.~W.}\ \bibnamefont {Sides}},\ }\bibfield
  {title} {\enquote {\bibinfo {title} {Metastable lifetimes in a kinetic Ising
  model: Dependence on field and system size},}\ }\href@noop {} {\bibfield
  {journal} {\bibinfo  {journal} {Phys.\ Rev.\ E}\ }\textbf {\bibinfo {volume}
  {49}},\ \bibinfo {pages} {5080} (\bibinfo {year} {1994})}\BibitemShut
  {NoStop}%
\bibitem [{\citenamefont {Gammaitoni}\ \emph {et~al.}(1998)\citenamefont
  {Gammaitoni}, \citenamefont {H{\"a}nggi},\ and\ \citenamefont
  {Jung}}]{GAMM98}%
  \BibitemOpen
  \bibfield  {author} {\bibinfo {author} {\bibfnamefont {L.}~\bibnamefont
  {Gammaitoni}}, \bibinfo {author} {\bibfnamefont {P.}~\bibnamefont
  {H{\"a}nggi}}, \ and\ \bibinfo {author} {\bibfnamefont {P.}~\bibnamefont
  {Jung}},\ }\bibfield  {title} {\enquote {\bibinfo {title} {Stochastic
  resonance},}\ }\href@noop {} {\bibfield  {journal} {\bibinfo  {journal}
  {Rev.\ Mod.\ Phys.}\ }\textbf {\bibinfo {volume} {70}},\ \bibinfo {pages}
  {223} (\bibinfo {year} {1998})}\BibitemShut {NoStop}%
\bibitem [{\citenamefont {Sides}\ \emph
  {et~al.}(1998{\natexlab{b}})\citenamefont {Sides}, \citenamefont {Rikvold},\
  and\ \citenamefont {Novotny}}]{SIDE98A}%
  \BibitemOpen
  \bibfield  {author} {\bibinfo {author} {\bibfnamefont {S.~W.}\ \bibnamefont
  {Sides}}, \bibinfo {author} {\bibfnamefont {P.~A.}\ \bibnamefont {Rikvold}},
  \ and\ \bibinfo {author} {\bibfnamefont {M.~A.}\ \bibnamefont {Novotny}},\
  }\bibfield  {title} {\enquote {\bibinfo {title} {Stochastic hysteresis and
  resonance in a kinetic Ising system},}\ }\href@noop {} {\bibfield  {journal}
  {\bibinfo  {journal} {Phys.\ Rev.\ E}\ }\textbf {\bibinfo {volume} {57}},\
  \bibinfo {pages} {6512} (\bibinfo {year}
  {1998}{\natexlab{b}})}\BibitemShut {NoStop}%
\bibitem [{\citenamefont {Korniss}\ \emph {et~al.}(2002)\citenamefont
  {Korniss}, \citenamefont {Rikvold},\ and\ \citenamefont {Novotny}}]{KORN02B}%
  \BibitemOpen
  \bibfield  {author} {\bibinfo {author} {\bibfnamefont {G.}~\bibnamefont
  {Korniss}}, \bibinfo {author} {\bibfnamefont {P.~A.}\ \bibnamefont
  {Rikvold}}, \ and\ \bibinfo {author} {\bibfnamefont {M.~A.}\ \bibnamefont
  {Novotny}},\ }\bibfield  {title} {\enquote {\bibinfo {title} {Absence of
  first-order transition and tri-critical point in the dynamic phase diagram of
  a spatially extended bistable system in an oscillating field},}\ }\href@noop
  {} {\bibfield  {journal} {\bibinfo  {journal} {Phys.\ Rev.\ E}\ }\textbf
  {\bibinfo {volume} {66}},\ \bibinfo {pages} {056127} (\bibinfo {year}
  {2002})}\BibitemShut {NoStop}%
\bibitem [{\citenamefont {Martin}(1977)}]{MART77}%
  \BibitemOpen
  \bibfield  {author} {\bibinfo {author} {\bibfnamefont {Ph.~A.}\ \bibnamefont
  {Martin}},\ }\bibfield  {title} {\enquote {\bibinfo {title} {On the
  stochastic dynamics of Ising models},}\ }\href@noop {} {\bibfield  {journal}
  {\bibinfo  {journal} {J.\ Stat.\ Phys.}\ }\textbf {\bibinfo {volume} {16}},\
  \bibinfo {pages} {149} (\bibinfo {year} {1977})}\BibitemShut {NoStop}%
%\bibitem [{\citenamefont {Vatansever}(2017)}]{VATA17}%
%  \BibitemOpen
%  \bibfield  {author} {\bibinfo {author} {\bibfnamefont {E.}~\bibnamefont
%  {Vatansever}},\ }\href@noop {} {\  (\bibinfo {year} {2017})},\ \bibinfo
%  {note} {private communication}\BibitemShut {NoStop}%
\bibitem{VATA17}
E.~Vatansever, ``Dynamically order-disorder transition in triangular lattice driven by a time dependent magnetic field," arXiv:1706.03351 (2017).
%
\bibitem [{\citenamefont {Kolmogorov}(1937)}]{KOLM37}%
  \BibitemOpen
  \bibfield  {author} {\bibinfo {author} {\bibfnamefont {A.~N.}\ \bibnamefont
  {Kolmogorov}},\ }\bibfield  {title} {\enquote {\bibinfo {title} {A
  statistical theory for the recrystallization of metals},}\ }\href@noop {}
  {\bibfield  {journal} {\bibinfo  {journal} {Bull.\ Acad.\ Sci.\ USSR, Phys.\
  Ser.}\ }\textbf {\bibinfo {volume} {1}},\ \bibinfo {pages} {355} (\bibinfo
  {year} {1937})}\BibitemShut {NoStop}%
\bibitem [{\citenamefont {Johnson}\ and\ \citenamefont {Mehl}(1939)}]{JOHN39}%
  \BibitemOpen
  \bibfield  {author} {\bibinfo {author} {\bibfnamefont {W.~A.}\ \bibnamefont
  {Johnson}}\ and\ \bibinfo {author} {\bibfnamefont {R.~F.}\ \bibnamefont
  {Mehl}},\ }\bibfield  {title} {\enquote {\bibinfo {title} {Reaction kinetics
  in processes of nucleation and growth},}\ }\href@noop {} {\bibfield
  {journal} {\bibinfo  {journal} {Trans.\ Am.\ Inst.\ Mining and Metallurgical
  Engineers}\ }\textbf {\bibinfo {volume} {135}},\ \bibinfo {pages} {416}
  (\bibinfo {year} {1939})}\BibitemShut {NoStop}%
\bibitem [{\citenamefont {Avrami}(1939)}]{AVRAMI}%
  \BibitemOpen
  \bibfield  {author} {\bibinfo {author} {\bibfnamefont {M.}~\bibnamefont
  {Avrami}},\ }\bibfield  {title} {\enquote {\bibinfo {title} {Kinetics of
  phase change},}\ }\href@noop {} {\bibfield  {journal} {\bibinfo  {journal}
  {J.\ Chem.\ Phys.}\ }\textbf {\bibinfo {volume} {7}},\ \bibinfo {pages}
  {1103} (\bibinfo {year} {1939})};\ \bibinfo {note} {{\bf 8}, 212 (1940);
  {\bf 9}, 177 (1941)}\BibitemShut {NoStop}%
\bibitem{RAMO99}
R.~A.\ Ramos, P.~A.\ Rikvold, and M.~A.\ Novotny, 
``Test of the Kolmogorov-Johnson-Mehl-Avrami picture of metastable decay
in a model with microscopic dynamics," Phys.\ Rev.\ B {\bf 59}, 9053 (1999). 
\bibitem [{\citenamefont {Binder}\ and\ \citenamefont {Virnau}(2016)}]{BIND16}%
  \BibitemOpen
  \bibfield  {author} {\bibinfo {author} {\bibfnamefont {K.}~\bibnamefont
  {Binder}}\ and\ \bibinfo {author} {\bibfnamefont {P.}~\bibnamefont
  {Virnau}},\ }\bibfield  {title} {\enquote {\bibinfo {title} {Overview:
  Understanding nucleation phenomena from simulations of lattice gas models},}\
  }\href@noop {} {\bibfield  {journal} {\bibinfo  {journal} {J.\ Chem.\ Phys.}\
  }\textbf {\bibinfo {volume} {145}},\ \bibinfo {pages} {211701} (\bibinfo
  {year} {2016})}\BibitemShut {NoStop}%
\bibitem{RIEG17B}
P.~Riego, P.~Vavassori, and A.~Berger, ``Towards an understanding of dynamic phase 
transitions," Physica B: Condensed Matter, in press 
https://doi.org/10.1016/j.physb.2017.09.043 (2017).
\bibitem [{\citenamefont {Langer}(1967)}]{LANG67}%
  \BibitemOpen
  \bibfield  {author} {\bibinfo {author} {\bibfnamefont {J.~S.}\ \bibnamefont
  {Langer}},\ }\bibfield  {title} {\enquote {\bibinfo {title} {Theory of the
  condensation point},}\ }\href@noop {} {\bibfield  {journal} {\bibinfo
  {journal} {Ann.\ Phys.\ (N.Y.)}\ }\textbf {\bibinfo {volume} {41}},\ \bibinfo
  {pages} {108} (\bibinfo {year} {1967})}\BibitemShut {NoStop}%
\bibitem [{\citenamefont {Langer}(1969)}]{LANG69}%
  \BibitemOpen
  \bibfield  {author} {\bibinfo {author} {\bibfnamefont {J.~S.}\ \bibnamefont
  {Langer}},\ }\bibfield  {title} {\enquote {\bibinfo {title} {Statistical
  theory of the decay of metastable states},}\ }\href@noop {} {\bibfield
  {journal} {\bibinfo  {journal} {Ann.\ Phys.\ (N.Y.)}\ }\textbf {\bibinfo
  {volume} {54}},\ \bibinfo {pages} {258} (\bibinfo {year} {1969})}\BibitemShut
  {NoStop}%
\bibitem [{\citenamefont {G{\"u}nther}\ \emph {et~al.}(1980)\citenamefont
  {G{\"u}nther}, \citenamefont {Nicole},\ and\ \citenamefont
  {Wallace}}]{GNW80}%
  \BibitemOpen
  \bibfield  {author} {\bibinfo {author} {\bibfnamefont {N.~J.}\ \bibnamefont
  {G{\"u}nther}}, \bibinfo {author} {\bibfnamefont {D.~A.}\ \bibnamefont
  {Nicole}}, \ and\ \bibinfo {author} {\bibfnamefont {D.~J.}\ \bibnamefont
  {Wallace}},\ }\bibfield  {title} {\enquote {\bibinfo {title} {Instantons and
  the {Ising} model below {$T_{\rm c}$}},}\ }\href@noop {} {\bibfield
  {journal} {\bibinfo  {journal} {J.\ Phys.\ A: Math.\ Gen.}\ }\textbf
  {\bibinfo {volume} {13}},\ \bibinfo {pages} {1755} (\bibinfo {year}
  {1980})}\BibitemShut {NoStop}%
\bibitem [{\citenamefont {Rikvold}\ and\ \citenamefont
  {Kolesik}(2000)}]{RIKV00B}%
  \BibitemOpen
  \bibfield  {author} {\bibinfo {author} {\bibfnamefont {P.~A.}\ \bibnamefont
  {Rikvold}}\ and\ \bibinfo {author} {\bibfnamefont {M.}~\bibnamefont
  {Kolesik}},\ }\bibfield  {title} {\enquote {\bibinfo {title} {Analytic
  approximations for the velocity of field-driven Ising interfaces},}\
  }\href@noop {} {\bibfield  {journal} {\bibinfo  {journal} {J.\ Stat.\ Phys.}\
  }\textbf {\bibinfo {volume} {100}},\ \bibinfo {pages} {377} (\bibinfo
  {year} {2000})}\BibitemShut {NoStop}%
\bibitem [{\citenamefont {Gunton}\ and\ \citenamefont {Droz}(1983)}]{GUNT83A}%
  \BibitemOpen
  \bibfield  {author} {\bibinfo {author} {\bibfnamefont {J.~D.}\ \bibnamefont
  {Gunton}}\ and\ \bibinfo {author} {\bibfnamefont {M.}~\bibnamefont {Droz}},\
  }\href@noop {} {\emph {\bibinfo {title} {Introduction to the Theory of
  Metastable and Unstable States}}}\ (\bibinfo  {publisher} {Springer-Verlag},\
  \bibinfo {address} {Berlin},\ \bibinfo {year} {1983})\BibitemShut {NoStop}%
\bibitem [{\citenamefont {Lifshitz}(1962)}]{LIFS62}%
  \BibitemOpen
  \bibfield  {author} {\bibinfo {author} {\bibfnamefont {I.~M.}\ \bibnamefont
  {Lifshitz}},\ }\bibfield  {title} {\enquote {\bibinfo {title} {Kinetics of
  ordering during second-order phase transitions},}\ }\href@noop {} {\bibfield
  {journal} {\bibinfo  {journal} {Sov.\ Phys.\ JETP}\ }\textbf {\bibinfo
  {volume} {15}},\ \bibinfo {pages} {939} (\bibinfo {year} {1962})}
  \bibinfo {note} {[Zh.\ {\'E}ksp.\ Teor.\ Fiz.\ {\bf 42}, 1354
  (1962)]}\BibitemShut {NoStop}%
\bibitem [{\citenamefont {Allen}\ and\ \citenamefont {Cahn}(1979)}]{ALLE79}%
  \BibitemOpen
  \bibfield  {author} {\bibinfo {author} {\bibfnamefont {S.~M.}\ \bibnamefont
  {Allen}}\ and\ \bibinfo {author} {\bibfnamefont {J.~W.}\ \bibnamefont
  {Cahn}},\ }\bibfield  {title} {\enquote {\bibinfo {title} {A microscopic
  theory for antiphase boundary motion and its application to antiphase domain
  coarsening},}\ }\href@noop {} {\bibfield  {journal} {\bibinfo  {journal}
  {Acta Metall.}\ }\textbf {\bibinfo {volume} {27}},\ \bibinfo {pages}
  {1085} (\bibinfo {year} {1979})}\BibitemShut {NoStop}%
\end{thebibliography}
%\bibliographystyle{prsty}
%\bibliographystyle{unsrt}
%\bibliographystyle{apsrev}

\clearpage
%=============================================


\clearpage

%\clearpage

\end{document}